

\input harvmac.tex
\def\pa{\partial}
\def\d{\delta}
\def\p{\pi}
\def\f{\phi}
\def\l{\lambda}
\def\o{\omega}
\def\ha{{1 \over 2}}
\def\frac#1#2{{#1\over#2}}
\def\W#1#2{$W_#1{}^{(#2)}$}

\Title{\vbox{\baselineskip12pt\hbox{LAVAL PHY-27/91}}}
{{\vbox {\centerline{On the classical $W_N{}^{(l)}$ algebras}}}}
\bigskip

\centerline{\it Didier A. Depireux and Pierre Mathieu}
\centerline{D\'epartement de Physique,}
\centerline{Universit\'e Laval}
\centerline{Qu\'ebec, Canada, G1K 7P4}
\medskip

\vskip.5in

\noindent
We analyze the \W Nl algebras according to their conjectured realization  as
the second Hamiltonian  structure of the integrable hierarchy resulting from
the interchange of $x$ and $t$ in the $l^{th}$ flow of the sl($N$) KdV
hierarchy. The \W 43 algebra is derived
explicitly along these lines, thus providing further support for the
conjecture.
This algebra is found to be equivalent to that obtained by the method of
Hamiltonian reduction. Furthermore, its twisted version reproduces the algebra
associated to a certain non-principal embedding of sl(2) into sl(4), or
equivalently, the u(2) quasi-superconformal algebra. General aspects of the \W
Nl algebras are also presented. We point out in particular that the $x
\leftrightarrow t$ interchange approach of the \W Nl algebra appears
straightforward only when $N$ and $l$ are coprime.

\vskip1in

\Date{Nov. 1991}
\newsec{Introduction}

Given a hierarchy of two-dimensional evolution equations, one can interchange
the role of the independent variables $x$ and $t$ for any member of the
hierarchy, thus producing a new integrable hierarchy of evolution equations.
Furthermore, such an interchange in two different equations of a given
hierarchy produces two new independent hierarchies. The idea of interchanging
the roles of $x$ and $t$ for integrable equations goes back to
 \ref\FO{B.Fuchssteiner and W.Oevel, Physica {\bf 145A} (1987) 67.}. There it
was
shown with simple examples (KdV, mKdV, sine-Gordon) that the resulting
hierarchy
is also integrable and bi-Hamiltonian.  For the KdV and mKdV cases, Kupershmidt
\ref\Kup{B.A.Kupershmidt, Phys.Lett. {\bf 156A} (1979) 53.} has obtained the
same
conclusion independently by considering a more general transformation (GL(2))
of
the independent variables. The Boussinesq equation in $x$--evolution has been
constructed in \ref\MO{P.Mathieu and W.Oevel, Mod. Phys. Lett.
{\bf A6} (1991) 2397.} and was shown to be bi-Hamiltonian and in fact
equivalent to the fractional KdV hierarchy of \ref\me{I. Bakas and D.A.
Depireux, Mod. Phys. Lett. {\bf A6}
(1991) 1561; Erratum {\it ibid.} {\bf A6} (1991) 2351.}.

These
statements have now been fully generalized in  \ref\Pi{M. de Groot, T.J.
Hollowood, J.L. Miramontes, ``{\it Generalized Drinfeld-Sokolov Hierarchies''},
IASSNS-HEP-91/19, PUPT-1251.}\ref\Pii{N.Burroughs, M.de Groot, T.J.Hollowood,
J.L.Miramontes,  ``{\it Generalized Drinfeld-Sokolov Hierarchies II: the
Hamiltonian structures''}, IASSNS-HEP-91/42, PUPT-1263;``{\it
Generalized $W$-algebras and Integrable Hierarchies''}, IASSNS-HEP-91/61,
PUPT-1285. }
 where various extensions of the Drinfeld-Sokolov \ref\DS{V.G. Drinfeld and
V.V. Sokolov, Sov. Math. Dokl. {\bf23} (1981) 457;  J. Sov. Math., {\bf 30}
(1985) 1975.}  approach to KdV-type equations have been worked out, including
hierarchies obtained by $x \leftrightarrow t$ interchange.

The interest for these new hierarchies obtained by $x \leftrightarrow t$
interchange is motivated by the potential conformal character of their second
Hamiltonian structure. Since the Hamiltonian character of an evolution equation
depends crucially upon which of the variables is chosen for the evolution, one
expects
a priori that the interchange of $x$ and $t$ will substantially modify the
Hamiltonian properties (i.e. the form of the Poisson brackets) of a given
integrable system.
 New conformal Poisson algebras would yield new
extended conformal algebras upon quantization, in the same way as the second
Hamiltonian structure of the usual
generalized KdV hierarchies are related to $W$-algebras
\ref\Wgeneral{P.Mathieu, Phys.Lett. {\bf B208} (1988) 101; I.Bakas, Phys.Lett.
{\bf B213} (1988) 313; Commun. Math. Phys. 123 (1989) 627; Y.Matsuo, Phys.Lett.
{\bf B227} (1989) 209; D.J. Smit, Commun. Math. Phys. {\bf128} (1990) 1;
A.Bilal
and J.L. Gervais, Phys.Lett. {\bf B206} (1988) 412; V.A.Fateev and
S.L.Lykyanov,
Int.J. Mod. Phys. {\bf A3} (1988) 507; J. Balog, L. Feher, P. Forgacs, L.
O'Raifeartaigh and A. Wipf, Ann.Phys. (N.Y.) {\bf 203} (1990) 76; P. Di
Francesco, C. Itzykson and J.-B. Zuber, Saclay SPhT 90-149.}. Recall that a
Hamiltonian structure is said to be conformal if it contains the Poisson
bracket
characterizing the second Hamiltonian  structure of the KdV equation, namely
$$\{u(x) ~,~ u(y) \} = ( \pa^3 + 4 u \pa + 2 u_x) ~ \d (x-y)~, \eqno(1.1)
$$
where
the fields on the RHS are evaluated at $x$ and $\pa \equiv \pa_x$.  The Fourier
transform of this bracket yields the Virasoro algebra realized in terms of
Poisson brackets \ref\Gervais{J.L.Gervais, Phys. Lett. {\bf B160} (1985) 277},
hence the name conformal. Actually, the classical \W 32
algebra of Polyakov \ref\Poly{A.M.Polyakov, Int.J. Mod. Phys. {\bf A5} (1990)
833; M.Bershadsky, Commun. Math. Phys. {\bf 139} (1991) 71.}, has been shown
\MO{} to be equivalent to the second Hamiltonian structure of the
Boussinesq (or sl(3)  KdV) hierarchy with $x$ and $t$
interchanged at the level of the Boussinesq equation itself.

It was conjectured in \MO{} that the second Hamiltonian structure of the
hierarchy obtained by interchanging $x$ and $t$ in the $l^{th}$ flow of the
sl($N$) KdV hierarchy (with $l < N$) would produce a new conformal algebra,
called \W N l (with \W N 1 $\equiv  W_{N}$). See also \Pi \Pii{} for similar
results and conjectures. We will call this new hierarchy the sl$(N)_l$
hierarchy.

It is simple to show that the restriction $l<N$ is necessary to produce a
conformal  \W Nl algebra. The sl($N$) KdV hierarchy is characterized by the
scalar Lax operator  $$L ~=~ \pa^N + u_2 \pa^{N-2} + \ldots + u_N~. \eqno(1.2)
$$
The evolution equations are
$$\pa_{t_l} L ~=~ [ (L^{l/N})_+ ~,~ L]~, \eqno(1.3) $$
where $+$ denotes the differential part of a pseudo-differential operator.
Interchanging $x$ and $t$ in the $l^{th}$ flow amounts to interchanging $t_l$
and
$t_1 = x $. In the newly produced hierarchy $t_l$ plays the role of the space
variable. In the normalization where dim$(x) = -1$, $t_l$ has dimensions $-l$.
To renormalize the dimensions of the new space variable to $-1$, one has to
divide all the dimensions by $l$. Thus the new algebra will contain bosonic
fields of fractional dimension (multiples of $1/l$). In order to be conformal,
it must contain a field of spin 2  (after dimensional renormalization). Now in
$(1.3)$, the evolution of the highest spin field takes the form
$$\pa_{t_l} u_N = c u_2{}^{(N+l-2)} + \ldots \quad,\quad c= {\rm constant},\
u^{(i)} = (\pa^i u)~. \eqno(1.4)
$$
As it will become clear below, the new set of independent fields required to
describe the system obtained by interchange of $t_1$ and $t_l$ can be chosen
generically to include $u_2,u_{2x},u_{2xx},\ldots,u_2{}^{(N+l-3)}$. Since
$u_2{}^{(i)}$ originally had dimension $2+i$, its new dimension will be
$(2+i)/l$. Thus the field $u_2{}^{(2l-2)}$ will have new dimension $2$, and it
belongs to the above sequence only if $l<N$.

Before pursuing the discussion of the second Hamiltonian structure of the
sl($N$) hierarchy, one should settle the question of its integrability. This is
most naturally discussed in terms of a zero curvature condition.
The usual sl($N$) hierarchy can be described by the scattering problem
$$\Phi_{t_l} = V^{(l)} \Phi \eqno(1.5)$$
where $\Phi = (\phi, \phi_x, \ldots)^T$, $V^{(l)} = (L)^{l/N}$ and $L\f =
\l\f$, $\l$ being the spectral parameter, i.e.
$$V^{(l)}{}_x - V^{(1)}_{t_l} + [
V^{(1)} , V^{(l)}] = 0 \eqno(1.6)
$$
with $x = t_1$. The interchange $t_1
\leftrightarrow t_l$ amounts to interchange the roles of $V^{(1)}$ and
$V^{(l)}$.
But the point is that we stick to a zero curvature formulation, hence this
operation manifestly preserves the integrability property (this is proved
rigorously in \Pi). In $V^{(l)}$ there is a constant piece, equal to
$$\Lambda^{(l)}{}_N = \l^i\,  \left(\matrix{ 0 & I_{N+j-1} \cr
               \l I_j & 0    \cr} \right) \qquad Ni+j=l \eqno(1.7)
$$
where $I_k$ is the $k\times k$ unit matrix. Hence for $l<N$, we find $i=0$ and
$j=l$ and the two matrices entering in $(1.6)$ are linear in $\l$.

One strategy to obtain the second Hamiltonian structure of the
sl$(N)_l$ KdV hierarchies is the following \MO. One first writes down the
$l^{th}$ flow of the sl$(N)$ KdV and mKdV hierarchies. These equations are
related by a Miura transformation  which characterizes the usual sl($N$)
hierarchy. The Miura map gives the free field representation of the classical
$W_N$ algebra, which is the second Hamiltonian structure of the sl($N$) KdV
hierarchy. Let us denote the corresponding Hamiltonian operator by $P_2$.
Similarly, denote by $\Theta$ the natural Hamiltonian operator of the sl($N$)
mKdV hierarchy\footnote{${}^1$}{It is also called the mKdV first Hamiltonian
structure for historical reasons (except in \DS). Also, it has lower
dimensions than the second one. Its naturalness is due to the fact that the
second Hamiltonian structure is both complicated and non-local.}. The canonical
character of the Miura map translates into the statement that  \ref\KW{B.A.
Kupershmidt and G.Wilson, Inv.Math. {\bf 62} (1981) 403.}\ref\Olver{See for
instance P.J.Olver, {\it Applications of Lie groups to differential equations},
Springer-Verlag, 1986.}
$$
P_2 = D\, \Theta D^{\dag} \eqno(1.8)
$$
where $D$ is the
Fr\'echet derivative of the KdV fields with respect to the modified KdV fields.
It is computed from the Miura map. $D^{\dag}$ is its formal adjoint. Thus the
$l^{th}$ flows of the sl($N$) KdV and mKdV hierarchies read as
$$\eqalign{u_{t_l} & = P_2   \nabla _u H_{\phantom m}~,\qquad
u=(u_2,u_3,\ldots,u_N)^T ~~,
\cr   p_{t_l} & = \Theta_{\phantom 2} \nabla_p
H_m ~,\qquad p = (p_1,p_2,\ldots,p_{N-1})^T~, \cr}  \eqno(1.9)
$$
where $\nabla_u = (\d/\d u_2, \ldots, \d/\d u_N)^T$ and similarly for
$\nabla_p$. $H$ is the appropriate Hamiltonian for the $l^{th}$ flow and $H_m$
is the expression of the
same Hamiltonian in terms of the modified fields. By interchanging $t_1$ and
$t_l$, one gets
$$\eqalign{
\tilde u_{t_l} & = \tilde P_2 \nabla _{\tilde u} \tilde H_{\phantom m}~, \cr
\tilde p_{t_l} & = \tilde \Theta_{\phantom2} \nabla_{\tilde p} \tilde H_m~,}
\eqno(1.10)
$$
where $\tilde u=(\tilde u_2,\tilde u_3,\ldots,\tilde u_{lN-l+1})^T$ and $\tilde
p
=  (\tilde p_1,\tilde p_2,\ldots,\tilde p_{lN-l})^T$ are the new independent
fields, whose number depends on $l$ (a canonical set of independent fields will
be displayed later). We want to find $\tilde P_2$. Notice that we know
$\tilde H$ since any conserved density $h$ for the sl($N$) hierarchy satisfies
$\pa_{t_l} h = \pa_x \tilde h$. Thus $\tilde h$ is a conserved density of the
new system. In (1.10), $\tilde H$ is the conservation law of appropriate
dimension. We also know the Miura map relating $\tilde p$ to $\tilde u$: it is
simply a rewriting of the usual Miura map where the $x$-derivatives of the
modified fields are eliminated by means of the $l^{th}$ sl($N$) mKdV equation.
Hence we also know $\tilde H_m$. Now as it will be illustrated below, there is
a
natural way to write the first equation of the sl$(N)_l$ mKdV
hierarchy (i.e. a choice of new modified fields $\tilde p$) which makes $\tilde
\Theta$ obtainable by inspection in a totally straightforward way. (It only
contains $\pa_t$ and constants, which makes its Hamiltonian character
manifest).
Having $\tilde \Theta$ and $\tilde D$,  the Fr\'echet derivative of $\tilde
u$ with respect to $\tilde p$, one can reconstruct $\tilde P_2$
 by
$$ \tilde P_2 = \tilde D\, \tilde \Theta \tilde D^{\dag}. \eqno(1.11)
$$
The Hamiltonian property of $\tilde P_2$ is thus inherited from that of
$\tilde \Theta$.
$\tilde P_2$ corresponds to the classical \W N l algebra.

The advantage of this construction, apart from its conceptual simplicity, is
that it gives directly the free field representation of the \W N l
algebra. A minor drawback is that $\tilde\Theta$ must be obtained by
inspection. Now the above procedure is totally straightforward in the
cases where $N$ and $l$ are coprime ($(N,l)=1$). However, when such is not the
case, the first sl$(N)_l$ equation appears under the form of a constrained
system (see e.g. (5.7)). We defer to another publication the detailed analysis
of
such cases, for which the simplest example is \W 4 2.\footnote{${}^2$}{\W 4 2
is
presented from the point of view of Hamiltonian reduction in \ref\CUNY{I.Bakas
and D.A.Depireux, ``{\it Self-Duality, KdV flows and $W$-algebras}'',
Proceedings
of the $XXth$ International Conference on Differential Geometric Methods in
Theoretical Physics, New-York, June 1991.}. After field redefinition and
twisting, it is equivalent to the particular $W$ algebra derived in
\ref\BTV{F.A. Bais, T. Tjin and P. van Driel, Nucl. Phys. {\bf B357} (1991)
632;
T. Tjin and P. van Driel, Amsterdam preprint, IFTA-91-04 (1991).} by
considering the embedding of sl(2) into sl(4) fixed by the decomposition
$\underline4 \rightarrow \underline2 + \underline2$ of the fundamental
representation.}

Here we work out in details a new  example of a \W N l algebra,
namely the \W 4 3 case.

This gives the first explicit form of a \W N l algebra for $N>3$, derived from
the point of view of integrable hierarchies. We will also check that the same
algebra can be obtained directly by the method of Hamiltonian reduction and the
corresponding flows can be extracted by reduction from sl(4) self-dual
Yang-Mills equations, as was the case for  \W 3 2 . After having worked out a
new non-trivial example of \W N l algebra, we will be in position to present a
set of general remarks concerning the structure of these algebras for
$(N,l)=1$,
including their spin content. But before, we illustrate the method by a simple
example.

\newsec{A simple example: Interchange of $x$ and $t$ for the usual KdV
equation}

Let us write the KdV equation under the form
$$u_t = u_{xxx} + 6 u u_x ~.\eqno(2.1)$$
Its two Hamiltonian structures are
$$\eqalignno{
u_t & = P_2 \nabla \ha \int u^2 dx & (2.2a)\cr
    & = P_1 \nabla \ha \int (2 u^3 - u_x{}^2) dx ~~,& (2.2b) }$$
with
$$P_2 = \pa^3 + 4 u \pa + 2 u_x \quad {\rm and} \quad P_1 = \pa~~, \eqno(2.3)
$$
The KdV equation in $x$-evolution is obtained as follows: one first introduces
two new independent fields
$$v = u_x \quad{\rm and}\quad w = u_{xx} \eqno(2.4)$$
so that (2.1) can be rewritten as
$$\left(\matrix{u\cr v \cr w \cr}\right)_x = \left(\matrix{ v \cr w \cr u_t -
6 u v \cr } \right) ~.\eqno(2.5) $$
This is the KdV equation in $x$-evolution or equivalently the first flow in the
sl$(2)_3$ hierarchy. One proceeds similarly with the mKdV equation
$$\eqalignno{
p_t = & p_{xxx} - 6 p^2 p_x                    & (2.6a) \cr
    = & \Theta_1 \nabla \ha \int (p_x^2 + p^4) & (2.6b) \cr
    = & \Theta_2 \nabla \ha \int p^2       ~,  & (2.6c) \cr  }$$
with
$$\Theta_1 = - \pa \quad {\rm and} \quad \Theta_2 = \pa^3 - 4 \pa p\
\pa^{-1} p \ \pa~. \eqno(2.7)$$
The Miura transformation
$$ u = p_x - p^2 \eqno(2.8) $$
is a canonical map from $\Theta_1$ to $P_2$. Indeed the Fr\'echet derivative of
$u$ with respect to $p$ is $\pa - 2 p$ so that
$D^{\dag} = - \pa - 2 p$ and \KW
$$P_2 = (\pa - 2 p ) ( - \pa) ( - \pa - 2 p)~. \eqno(2.9)$$
Introducing
$$ q = p_x,~r = p_{xx}~, \eqno(2.10) $$
one can rewrite the mKdV equations in the form
$$\left(\matrix{ p \cr q \cr r \cr } \right)_x =
\left(\matrix{ q \cr r \cr p_t + 6 p^2 q \cr } \right)~. \eqno(2.11) $$
Now in order to find $\tilde P_2$, the second Hamiltonian structure for (2.5),
we will need $\tilde\Theta_1$, the first Hamiltonian structure for (2.11). As
already pointed out this must be found by inspection. We now show that with a
simple field redefinition, this step is straightforward. The trick is to look
for the field transformation which simplifies maximally the equation of the
highest degree
in (2.11). Here this amounts to introducing a new variable $s$ linearly related
to $r$ such that $s_x = p_t$. Thus we choose $s = r - 2 p^3$ and (2.11) becomes
$$\left(\matrix{
p \cr q \cr s \cr } \right)_x =
\left(\matrix{ q \cr s + 2 p^3 \cr p_t \cr } \right)~. \eqno(2.12)
$$
We want to write the RHS under the form $\tilde\Theta_1\nabla \tilde
H_m$. $\tilde H_m$ is the $x \leftrightarrow t$ interchanged version of
$\int (p^4 + p_x^2) dx$, that is $\int (p^4+\ldots) dt$. Hence we look for the
density $(p^4 + \ldots)$ such that $\pa_x(p^4+\ldots) = \pa_t(\ldots)$, where
in
the RHS one has a usual mKdV conserved density. Dimensionally one sees that it
is $p^2$. Since $\pa_t p^2 = \pa_x(2 p p_{xx} - p_x^2 - 3 p^4)$, and
$\tilde H_m$ is only defined up to a multiplicative constant, we choose
$\tilde H_m$ as
$$\tilde H_m = \int (\frac 32 p^4 + \ha p_x^2 - p p_{xx}) dt = \int (- \ha p^4
+ \ha q^2 - ps)dt. \eqno(2.13)$$
Of course, using (2.12),  it is simple to check explicitly that
$(\tilde H_m)_x=0$. An even more direct approach is the following: we know that
the KdV conservation laws can be obtained as $\int {\rm Res} L^{k/2} dx$ with
$L
= \pa^2 + u$.  Writing
${\rm Res} L^{k/2}$ in terms of the new fields, one gets directly the new
conserved densities. For example, ${\rm Res}L^{3/2} = \frac18 ( u^2 + \frac
13 u_{xx}) $ so that one can take $\tilde H$ to be $\frac32 \int (u^2 + \frac13
w) dt$ which  gives directly the above $\tilde H_m$, using the Miura
transformation presented below in (2.16).
 We now search for a matrix differential operator $\tilde\Theta_1$ such that
$$\left(\matrix{
q         \cr
s + 2 p^3 \cr
p_t       \cr } \right) = \tilde \Theta_1
\left(\matrix{ - s - 2 p^3 \cr q \cr - p \cr } \right)~~. \eqno(2.14) $$
The elements of $\tilde\Theta_1$ are easily found by inspection to be
$$\tilde\Theta_1 = \left(\matrix{
0 & 1 & 0       \cr
-1 & 0 & 0      \cr
0 & 0 & - \pa_t \cr }\right)~~. \eqno(2.15)$$
We claim that once the fields are chosen such that the highest degree modified
equation has the form $\phi_x = \psi_t$, $\tilde\Theta_1$ is always obtained as
simply as above. Now let us work out the Miura transformation:
$$ \eqalign{
u =& p_x - p^2 = q - p^2   ~~,                     \cr
v =& u_x = p_{xx} - 2 p p_x = s + 2 p^3 - 2 p q ~~,\cr
w =& u_{xx} = p_{xxx} - 2 p_x^2 - 2 p p_{xx} = p_t - 2 q^2 - 2 p s - 4 p^4 + 6
p^2 q~.\cr}\eqno(2.16)$$
The Fr\'echet derivative of $(u,v,w)^T$ with respect to $(p,q,s)^T$ is found to
be
$$\tilde D = \left(\matrix{
- 2 p & 1 & 0\cr
6 p^2 - 2 q & - 2 p & 1 \cr
\pa_t - 2 s - 16 p^3 + 12 p q & - 4 q + 6 p^2 & - 2 p \cr}\right) \eqno(2.17)$$
and
$$\tilde D^{\dag} = \left(\matrix{
- 2 p & 6 p^2 - 2 q & -\pa_t - 2 s - 16 p^3 + 12 p q \cr
1     & - 2 p       &  - 4 q + 6 p^2                 \cr
0     & 1           & - 2 p                          \cr}\right)~~.
\eqno(2.18)$$

Now it is simple to obtain $\tilde P_2$, from the matrix product
(1.11). The result is \FO\Kup\Pii
$$\tilde P_2 = \pmatrix{0 & 2 u & \pa_t + 2 v \cr
- 2 u & - \pa_t & 2 w + 12 u^2 \cr
\pa_t - 2 v & - 2 w - 12 u^2 & - 8 u \pa_t - 4 w \cr}~~. \eqno(2.19)
$$
The first Hamiltonian structure can be obtained by shifting $u$ by a constant
factor, i.e. with $u \rightarrow u+\l$, $\tilde P_2 \rightarrow \tilde P_2 +
2 \l \tilde P_1$ where
$$
\tilde P_1 = \pmatrix{0 & 1 & 0 \cr
-1 & 0 & 12 u \cr
0 & - 12 u  & - 4 \pa_t \cr}~~. \eqno(2.20)
$$
The field redefinition $w\rightarrow w+ 6 u^2$ transforms $\tilde P_1$ into a
manifestly Hamiltonian form similar to (2.15). Notice that to recover the first
Hamiltonian structure, one should not shift the field of highest spin in the
new
set of independent fields, but merely the highest spin field of the original
set. On the other hand,  given $\tilde\Theta_1$, one can calculate
$\tilde\Theta_2$ as follows \FO: the master-symmetries for (2.12) can be
obtained directly from those of the mKdV  equation with $x$ and $t$
interchanged. Let $T$ be the analogue, for (2.12), of the first time dependent
symmetry for the mKdV equation. Then $\tilde\Theta_2$ is, up to a
multiplicative
factor, the Lie derivative of $\tilde\Theta_1$ with respect to $T$.

Thus the operator (2.19) characterizes the second Hamiltonian structure of the
sl$(2)_3$ KdV hierarchy. This operator clearly does not define a $W$-algebra,
since it is not conformal. Indeed, by rescaling the dimensions such that
dim$(\pa_t) = 1$, so that one must divide the dimension of all the fields by 3,
one gets dim$(u,v,w) = 2/3,1,4/3$. Hence there is no spin-2 field.

As a final
remark, we stress that {\it all\/} the equations of the sl$(2)_3$ hierarchy can
be obtained systematically from those of the sl(2) hierarchy by interchanging
$t_l = t_3$ and $t_1 = x$ at the level of the KdV equation itself. For
instance,
the $j^{th}$ flow in the ordinary KdV hierarchy takes the form
$$u_{t_j} = f_j(u,u_x,u_{xx},u_{xxx},\ldots) \eqno(2.21)$$
The $x$-derivatives are eliminated by means of (2.4) and (2.1) with the
result that (2.2) is transformed into
$$u_{t_j} ~=~ g_j(u,v,w,u_{t_3},v_{t_3},w_{t_3},...) \eqno(2.22)$$
and similar expressions for $v_{t_j}$, $w_{t_j}$.

\newsec{The classical $W_4{}^{(3)}$ algebra by $x \leftrightarrow t$
interchange.}

\subsec{Generalities on the second Hamiltonian structure of scalar Lax
equations.}

Introduce the pseudo-differential operator
$$F^{(l)} = \sum_{k=1}^N \pa^{-k} f_{N-k+1}^{(l)}\eqno(3.1)$$
where $f_1^{(l)}$ is fixed by the condition
$${\rm Res}~[F^{(l)},L]=0 ~,\eqno(3.2)$$
with $L$ the scalar Lax operator (1.1) and ${\rm Res} \sum_i a_i \pa^i =
a_{-1}$. The second Hamiltonian structure of the sl$(N)$ KdV hierarchy takes
the
form \KW\DS
$$\pa_{t_l} L = (L F^{(l)})_+ L - L ( F^{(l)} L)_+~, \eqno(3.3)$$
which translates into
$$(u_i)_{t_l} = (P_2)_{ij} f_j^{(l)}~.\eqno(3.4)$$
$P_2$ gives the Poisson brackets of the different fields, i.e.
$$\{ u_i(x),u_j(y) \} = (P_2(x))_{ij} \d(x-y) ~.\eqno(3.5)$$
The Miura tranformation, which furnished the free field realization of this
Poisson algebra, can be obtained as follows: one first factorizes $L$ as
$$L = (\pa + \f_{N-1})(\pa+\f_{N-2})\ldots(\pa+\f_1)(\pa+\f_0) ~,\eqno(3.6)$$
where $\sum_0^{N-1} \f_i = 0 $. The $\f_i$'s are then expressed in terms of a
set of linearly independent fields $p_i$'s by
$$\f_k = \sum_{i=1}^{N-1} \o^{ki} p_i \quad,\quad \o = e^{2i\p/N}~. \eqno(3.7)
$$
The Poisson structure for the $p_i$'s is \KW \DS
$$ \{ p_i(x) , p_j(y) \}  = - {1\over N} \d_{N-i,j} \d_x(x-y)~. \eqno(3.8) $$
These brackets can be diagonalized, by introducing the fields
$$r_j = \left\{ \eqalign{
& \phantom{-i}\sqrt{N\over 2}   ( p_j + p_{N-j})~, \quad j<{N\over 2} \cr
& \phantom{-i}\sqrt{N} ~\phantom{(}p_j \phantom{+ p_{N-j})}~~, \quad j={N\over
2}
\cr
&          -i \sqrt{N\over 2}   ( p_j - p_{N-j})~, \quad j>{N\over 2} \cr}
\right. \eqno(3.9)$$
so that
$$\{ r_i(x),r_j(y) \} = - \d_{ij}  \d_x(x-y)~.\eqno(3.10)$$

\subsec{Specialization to the sl(4) case.}

The sl(4) scalar Lax operator is
$$L = \pa^4 + u_2 \pa^2 + u_3 \pa + u_4~. \eqno(3.11)$$
The components $(i,j)$ of the corresponding operator $P_2$ are then found to be
$$ \eqalign{
(2,2):\quad &   5 \pa^3 + u_2 \pa + \pa u_2\cr
(2,3):\quad & - 5 \pa^4 - 2 \pa^2 u_2 + 2 \pa u_3 + u_3 \pa \cr
(2,4):\quad & {3\over 2} \pa^5 + {3\over 2} \pa^3 u_2 - {3\over 2} \pa^2 u_3 +
3 \pa u_4 + u_4 \pa \cr
(3,3):\quad & - 6 \pa^5 - 2 ( \pa^3 u_2 + u_2 \pa^3 ) + (\pa^2 u_3 - u_3 \pa^2)
+ 2 (u_4 \pa + \pa u_4) - \cr
& \quad\ha (u_2 \pa + \pa u_2) \cr
(3,4):\quad & 2 \pa^6 + 2 \pa^4 u_2 + {3 \over 2} u_2 \pa^4 - 2 \pa^3 u_3  + 3
\pa^2 u_4 - u_4 \pa^2 + \ha u_2 \pa^2 u_2 - \ha u_2 \pa u_3 \cr
(4,4):\quad & {3\over 4} \pa^7 + {3\over 4}(u_2 \pa^5 + \pa^5 u_2) + {3\over
4} (u_3 \pa^4 - \pa^4 u_3) + (u_4 \pa^3 + \pa^3 u_4 ) + \cr
& \quad{3\over 4} u_2 \pa^3
u_2 + u_2 u_4 \pa + \pa u_2 u_4 - {3\over 4} u_3 \pa u_3 + {3\over 4} ( u_3
\pa^2
u_2 - u_2 \pa^2 u_3)\cr}
\eqno(3.12) $$
Now by rewriting $L$ under the form
$$L= (\pa + \f_{3})(\pa+\f_{2})(\pa+\f_1)(\pa+\f_0)~, \eqno(3.13)$$
one can express the fields $u_i$ in terms of the $\f_i$'s and ultimately using
(3.7) and (3.9), in terms of the fields $r_i$'s. The result turns out to be
$$
\eqalignno{
u_2 = & - \ha ( r_1^2 + r_2 ^2 + r_3 ^2 ) + {\sqrt{2}} r_{1x} + r_{2x} -
{\sqrt{2}}
r_{3x}~,\cr
u_3 = & ~\frac 3 {\sqrt{2}} r_{1xx} + r_{2xx} - {1\over {\sqrt{2}}}
 r_{3xx}
- \frac32 r_1 r_{1x} - \frac 1{\sqrt{2}}  r_{1x} r_2 + \ha r_{1x} r_3 - \frac
1{\sqrt{2}} r_1 r_{2x} - \cr
{} & \quad r_2 r_{2x} - \frac 1{\sqrt{6}} r_{2x} r_3 - \ha r_1
r_{3x} - \frac 1{\sqrt{2}} r_2 r_{3x}
- \ha r_3 r_{3x} + \ha r_1^2 r_2 - \ha r_2 r_3^2~, \cr
u_4 = & ~\frac 1{\sqrt{2}} r_{1xxx} + \ha r_{2xxx} - \ha r_1 r_{1xx} - \frac
1{2\sqrt{2}} r_2 r_{1xx} - \frac 1{\sqrt{2}} r_1 r_{2xx} - \cr
{} & \quad \ha r_2 r_{2xx} - \ha r_1 r_{3xx} - \frac 1{2 \sqrt{2}} r_2 r_{3xx}
+
\frac 34 r_1 r_{1x} r_2 + \cr
{} & \quad \frac1{2 \sqrt{2}} r_1 r_{1x} r_3 + \frac14 r_{1x} r_2 r_3 -
\frac 1{2 \sqrt{2}} r_{1x} r_3^2 + \frac14 r_1^2 r_{2x} + \frac1{2 \sqrt{2}}
r_1
r_2 r_{2x} - \cr
{} & \quad \frac 14 r_2^2 r_{2x} - \ha r_1 r_{2x} r_3 - \frac 1{2\sqrt{2}} r_2
r_{2x} r_3 - \frac14 r_{2x} r_3^2 + \frac 1{2 \sqrt{2}} r_1^2 r_{3x} + \cr
{} & \quad \frac14 r_1 r_2 r_{3x} - \frac1{2 \sqrt{2}} r_1 r_3 r_{3x} - \frac
14
r_2 r_3 r_{3x} - \ha r_{1x} r_{1x} - \frac 1{\sqrt{2}} r_{1x} r_{2x} - \cr
{} & \quad \frac 14
r_{2x} r_{2x} - r_{1x} r_{3x} - \frac1{\sqrt{2}} r_{2x} r_{3x} - \frac 18 r_1^2
r_2^2 + \frac 1{16} r_2^4 + \frac 14 r_1^2 r_3^2 - \frac 18 r_2^2 r_3^2.
 & (3.14) \cr } $$
The first Hamiltonian structure for these modified fields is given by the
Hamiltonian operator $\Theta$ of (3.10) (we omit the subscript 1)
$$
\Theta = \left(\matrix{ - \pa & 0 & 0 \cr 0 & - \pa & 0 \cr 0 & 0 & - \pa
\cr} \right) \eqno(3.15)
$$
Again one can check explicitly the canonical character of the above Miura map
by checking directly the identity $P_2 = D \Theta D^{\dag}$.

In the following we will be interested more particularly in the third flow  of
the sl(4) KdV hierarchy. It can be computed from
$$
\eqalign{ L_t =&  [ (L^{3/4})_+ , L ]\qquad {\rm where\ }t=t_3 \cr
\noalign{\hbox{ and}}
\ L^{3/4} =& \pa^3 + \frac 34 u_2 \pa + (\frac 34 u_3 -
\frac 38 u_{2x} ) + \cr
& \quad \frac34 ( u_4 - \ha u_{3x} + \frac1{12} u_{2xx} - \frac18
u_2{}^2 ) \pa^{-1} + \ldots\cr} \eqno(3.16) $$
One obtains
$$
\eqalign{
u_{2t} = & \frac 14 u_{2xxx} - \frac 32 u_{3xx} + 3 u_{4x}  - \frac 34 u_2
u_{2x},  \cr
u_{3t} = & \frac 34 u_{2xxxx} - 2 u_{3xxx} + 3 u_{4xx} - \frac 34 u_2 u_{3x} -
\frac 34 u_{2x} u_3, \cr
u_{4t} = & \frac 38 u_{2xxxxx} - \frac 34 u_{3xxxx} + u_{4xxx} + \frac 38 u_2
u_{2xxx} - \frac 34 u_2 u_{3xx} +  \cr
& \quad \frac 38 u_{2xx} u_3 + \frac 34 u_2 u_{4x} - \frac 34 u_3 u_{3x}. \cr}
\eqno(3.17) $$
These equations have the following Hamiltonian formulation:
$$\left(\matrix{
u_2 \cr u_3 \cr u_4 \cr}
\right)_t = P_2 \nabla _u \int (u_4 - \frac 18 u_2^2) dx ~,
$$
the Hamiltonian density being ${\rm Res}\ L^{3/4}$ up to total derivatives. We
will also be interested in the modified version of this equation. Rewriting the
above Hamiltonian in terms of the modified fields, one has
$$\eqalign{
H_m = \int [-\frac 14 & r_{1x} r_{1x} + \frac 18 r_{2x} r_{2x} - \frac 14
r_{3x}
r_{3x}  - \cr
{} & \frac 34 r_1 r_{2x} r_3 - \frac 1{32} (r_1^4 + 6 r_1^2 r_2^2 - 6
r_1^2 r_3^2 - r_2^4 + 6 r_2^2 r_3^2 + r_3^4)]~,\cr} \eqno(3.18)
$$
and the corresponding mKdV equations are
$$\left(\matrix{ r_1 \cr\noalign{\vskip3pt} r_2 \cr\noalign{\vskip3pt} r_3 \cr
}
\right)_t = -
\left( \matrix{ \ha r_{1xx} - \frac 34 r_{2x} r_3 - \frac 18 r_1^3 - \frac 38
r_1 r_2^2 + \frac 38 r_1 r_3^2 \cr \noalign{\vskip3pt}
- \frac 14 r_{2xx} + \frac 34 (r_1 r_3)_x - \frac 38 r_1^2 r_2 + \frac 18 r_2^3
- \frac 38 r_2 r_3^2 \cr \noalign{\vskip3pt}
\ha r_{3xx} - \frac 34 r_1 r_{2x} + \frac 38 r_1^2 r_3 - \frac 38 r_2^2 r_3 -
\frac 18 r_3^3\cr } \right)_x ~~.\eqno(3.19)
$$
\vskip8pt

\subsec{The Hamiltonian structure of the sl$(4)_3$ hierarchy.}

We now want to rewrite (3.17) and (3.19) as evolution equations with respect to
$x$. Let us start with (3.17), and introduce the fields
$$ u_{ix} = v_i~~,~~u_{ixx} = w_i~~,~~~i=2,3,4~. \eqno(3.20)$$
One finds
$$\eqalign{
 u_{ix} =& v_i~,~~v_{ix} = w_i~,\cr
w_{2x} = & 4 u_{2t} + 6 w_3 - 12 v_4 + 3 u_2 v_2~, \cr
w_{3x} = & - \frac65 v_{2t} + \frac25 u_{3t} + \frac{12}5 w_4 - \frac9{10}
v_2{}^2 - \frac9{10} u_2 w_2 + \frac 3{10} u_2 v_3 + \frac3{10} v_2 u_3~, \cr
w_{4x} = & 3 w_{2t} - 6 v_{3t} + 10 u_{4t} - 6 u_2 u_{2t} - 6 u_2 w_3 +
\frac{21}2 u_2 v_4 - \cr
& \quad \frac92 u_2{}^2 v_2 + \frac{27}4 v_2 w_2 - 9 v_2 v_3 + \frac{15}2 u_3
v_3 - \frac{33}4 w_2 u_3 ~. \cr } \eqno(3.21)
$$
The $x \leftrightarrow t$ interchange version of $H_m$ read off from
${\rm Res} L^{3/4} $ in (3.16) and reexpressed in terms of the above fields is
$$ \tilde H  = \int ( u_4 - \frac18 u_2{}^2 - \ha v_3 + \frac1{12} w_2) dt~,
\eqno(3.22) $$
and we are looking for the corresponding Hamiltonian operator $\tilde P_2$
which
allows the rewriting of the above system in the form (1.10). For this we need
first the modified version of (3.21) and its natural Hamiltonian structure.
To write (3.19) in $x$-evolution we introduce the variables $r_{ix} = s_i$,
$r_{ixx} = \tilde q_i$. However, using the hindsight gained from the study of
the KdV and the Boussinesq cases, we choose new variables $q_i$ so as to keep
the highest field equations in the form $q_{ix} = r_{it}$. The explicit form of
the $q_i$'s can then be read off directly from (3.19), i.e. $q_1 = - \ha
r_{1xx} + \frac34 r_{2x} r_3 + \ldots$, and the sl$(4)_3$ version of (3.19) is

$$ \eqalignno{
{} & \left\{ \eqalign{
r_{1x} &= s_1 \cr
s_{1x} &= - 2 q_1 - 2[ - \frac 34 s_2 r_3 - \frac 18 r_1^3 - \frac 38 r_1 r_2^2
+ \frac 38 r_1 r_3^2]\cr
q_{1x} &= r_{1t} \cr } \right.& (3.23a) \cr
{} & \left\{ \eqalign{
r_{2x} &= s_2 \cr
s_{2x} &= 4 q_2 + 4[ \frac 34 s_1 r_3 + \frac 34 r_1 s_3 - \frac 38 r_1^2 r_2
+ \frac 18 r_2^3 - \frac 38 r_2 r_3^2]\cr
q_{2x} &= r_{2t} \cr } \right. & (3.23b) \cr
{} & \left\{ \eqalign{
r_{3x} &= s_3 \cr
s_{3x} &= - 2 q_3 - 2[ - \frac 34 r_1 s_2 + \frac 38 r_1^2 r_3 - \frac 38 r_2^2
r_3 - \frac 18  r_3^3]\cr
q_{3x} &= r_{3t}\cr } \right.& (3.23c) \cr}
$$

These flows can be written in a Hamiltonian form as follows. The Hamiltonian
can
be easily obtained from the above $\tilde H$, where we express the KdV fields
in terms of the modified fields,  using the Miura map (3.14) and
the equations (3.23) to eliminate the $x$-derivative of the modified KdV
fields. We get
$$\eqalign{
\tilde H_m = \int & [r_1 q_1 + r_2 q_2 + r_3 q_3 + \frac14
s_1^2 - \frac18 s_2^2 + \frac14 s_3^2 - \cr
{} & \quad \frac 1{32} r_1^4 - \frac 3{16} r_1^2 r_2^2 + \frac 3{16} r_1^2
r_3^2 +
\frac 1{32} r_2^4 - \frac 3{16} r_2^2 r_3^2 - \frac 1{32}
r_3^4]\,dt\cr}\eqno(3.24)
$$
We determine
$\tilde\Theta$ by inspection, writing $(r,s,q)^T_x = \tilde \Theta(\nabla_r
\tilde H_m, \nabla_s \tilde H_m,\nabla_q \tilde H_m)^T$, with the result:
$$
\tilde \Theta = \bordermatrix{
& r & s & q \cr
r & {\bf 0} & \matrix{ 2 & {} & {} \cr {} & -4 & {} \cr {} & {} & 2 \cr} &
{\bf 0} \cr
s & \matrix{ -2 & {} & {} \cr {} & 4 & {} \cr {} & {} & -2 \cr} &
\matrix{ {} & - 6 r_3 & {} \cr 6 r_3 & {} & 6 r_1 \cr {} & - 6 r_1 & {} \cr} &
{\bf 0} \cr
q & {\bf 0} & {\bf 0} & \matrix{ \pa_t & {} & {} \cr {} & \pa_t & {} \cr
{} & {} & \pa_t
\cr} \cr}~. \eqno(3.25)
$$

This operator can be further simplified by introducing the new fields
$\tilde s_i = s_i - 3 \d_{2i} r_1 r_3$, $2 \tilde r_1 = r_1$,
$- 4 \tilde r_2 = r_2$, $2 \tilde r_3 = r_3$, so that $\tilde\Theta$ takes the
form
$$
\tilde \Theta = \bordermatrix{
& \tilde r & \tilde s & q \cr
\tilde r & {\bf 0}      & {\bf I}_3 &  {\bf 0}         \cr
\tilde s & - {\bf I}_3  & {\bf 0}   & {\bf 0}          \cr
q        &   {\bf 0}    & {\bf 0}   & {\bf I}_3 \pa_t  \cr} ~~.\eqno(3.26)
$$
Since this is antisymmetric and field independent (so that the Jacobi
identities are automatically satisfied), this operator is manifestly
Hamiltonian. Now, having $\tilde\Theta$ and the Miura map, which leads to
$\tilde D$,
one can calculate $\tilde P_2$ by (1.11). The explicit form of $\tilde P_2$
obtained this way is not manifestly conformal. However, after some field
redefinitions given in the next section, it can be transformed into a conformal
algebra, to be presented in that same section. Exactly the same algebra can be
derived by the method of Hamiltonian reduction to which we now turn.

\newsec{The classical \W 4 3 algebra by the method of Hamiltonian reduction.}

\subsec{\W 4 3 by Hamiltonian reduction.}

In this section, we derive the \W43 algebra by the method of Hamiltonian
reduction. This method is by now standard and
will not be reviewed here \ref\YanLect{For a good review, see I.Bakas,``{\it
Self-Duality, Integrable Systems, $W$-Algebras and all that}'', $27^{th}$
Karpacz Winter School of Theoretical Physics, February 1991.}. We
start from a 1-dim connection that depends on a coordinate $t$, taking values
in
the algebras of sl(4). In the matrix representation we constrain the $(1,4)$
element of the connection to be -1, i.e. $$A(t) = \pmatrix{
J_{11} & J_{12} & J_{13}    & -1 \cr
J_{21} & J_{22}  & J_{23}   & J_{24}   \cr
J_{31} & J_{32}  & J_{33}   & J_{34}   \cr
J_{41} & J_{42}  & J_{43}   & J_{44}   \cr } ~,\quad J_{44} = -
\sum_{i=1}^3 J_{ii}~.\eqno(4.1) $$
The form of the
constraint is preserved by the gauge transformations
$$
g^{-1} (\pa_t + A(t)) g =
\pa_t + A^g(t)~~, \eqno(4.2)
$$
where $g$ is a lower triangular matrix with 1's on
the diagonal. We cannot completely fix this gauge invariance in a local way.
However, restricting ourselves to the subalgebra with $g_{23} = 0 $, we can
fix the gauge invariance with respect to this subalgebra; using (4.2),
we bring $A(t)$ to the canonical form
$$Q(t) = \pmatrix{
0         & 0        & 0        & -1         \cr
A_1\quad  & U_1\quad & Z\quad   & 0          \cr
B_1       & A_2      & U_2      & 0          \cr
T         & B_2      & A_3      & -U_1-U_2   \cr }~~. \eqno(4.3)
$$
 Using the
expression of the fields appearing in $Q$ in terms of the original currents of
$A(t)$ (which form an sl($N$) current algebra), we obtain an algebra which we
call the \W43 algebra. The superscript 3 in this context corresponds to the
fact
that the third upper diagonal has been set to -1. The spin
content of the algebra can be read off directly from (4.3) since the dimensions
are constant along the diagonals and by moving from the top right corner to the
bottom left corner, they increase by units of $1/3$. Since a constant has
dimension 0, we find ${\rm dim}(Z,U_i,A_i,B_i,T) = (2/3,1,4/3,5/3,2)$.

 After introducing $T_0 = T - A_2 Z - U_1^2 -
U_2^2 - U_1 U_2 + \frac 23 U_{1t} + \frac 13 U_{2t}$, which makes $T_0$ into a
classical energy-momentum tensor, we  get the algebra $$\eqalignno{
&\{Z,U_1\} = Z \d \quad,\quad \{Z,U_2\} = -Z \d \quad,\quad \{Z,A_2\} = - \d_t
+
(U_2 - U_1) \d \cr
& \{ Z,B_1\} = - A_1 \d ~,~
\{ Z , B_2 \} = A_3 \d ~,~ \{ U_1, U_1\} = -\frac34 \d_t \quad,\quad \{
U_1,U_2 \} = \frac14 \d_t \cr
& \{ U_1,A_1 \} = - A_1 \d ~,~
\{ U_1, A_2 \} = A_2 \d ~,~ \{ U_1 , B_2 \} = B_2 \d ~,~ \{
U_2,U_2 \} = - \frac34 \d_t \cr
& \{ U_2,A_2 \} = - A_2 \d ~,~ \{ U_2,A_3 \} = A_3 \d ~,~ \{
U_2,B_1 \} = - B_1 \d ~,~ \{ A_1,A_2 \} = B_1 \d \cr
& \{ A_1, A_3 \} = 2 Z \d_t + ( Z_t + 2 (U_1 + U_2) Z ) \d \quad,\quad
\{ A_2,A_3 \} = B_2 \d  \cr
& \{ A_1 , B_2 \} = \d_{tt} + (3 U_1 + U_2 ) \d_t + \cr
& \qquad\qquad ( \frac43 U_{1t} + \frac23 U_{2t} + T_0 + 2 Z A_2 + 3 U_1^2 + 2
U_1 U_2 + U_2^2) \d \cr
& \{ A_3 , B_1 \} = - \d_{tt} + (U_1 + 3 U_2 ) \d_t + \cr
& \qquad\qquad ( \frac23 U_{1t} + \frac43 U_{2t} - T_0 - 2 Z A_2 - U_1^2 - 2
U_1 U_2 - 3 U_2^2) \d \cr
& \{ B_1, B_2 \} = 2 A_2 \d_t + ( A_{2t} + 2 ( U_1 + U_2)A_2) \d\cr
& \{ T_0, Z\} = \frac23 Z \d_t - \frac13 Z_t \d \cr
& \{ T_0, U_1\} = - \frac16 \d_{tt} + U_1\d_t \quad,\quad     \{ T_0, U_2\} =
    \frac16 \d_{tt} + U_2\d_t  \cr
& \{ T_0, A_i\} = \frac43 A_i \d_t + \frac13
A_{it}\d \cr & \{ T_0, B_i\} = \frac53 B_i \d_t + \frac23 B_{it}\d \cr
& \{ T_0, T_0\} = \frac 59 \d_{ttt} + 2 T_0 \d_t + T_{0t} \d &(4.4)\cr }$$
All other brackets vanish.
Here the two fields in the Poisson brackets are evaluated at $t$ and $t'$
respectively; all the fields on the RHS are evaluated at $t$ and $\d =
\d(t-t')$. We have checked that the Jacobi identities are satisfied for this
algebra.

\subsec{A twisted version of the \W 43 algebra and the relation to covariantly
coupled algebras and quasi-superconformal algebras.}

Note the $\{T_0,U_i\}$ relations in (4.4), which display the non-primary
character of the $U_i$ fields. (Recall that a field $\f$ is primary, with
dimension $h$ if it satisfies  $\{T_0,\f\} = h \f \d_t + (h-1) \f \d$).
This could be cured by taking $T_N = T_0 - \frac16 U_{1t} + \frac16 U_{2t}$ as
the energy-momentum tensor. Then all the fields are primary, but their spins,
as
read off their Poisson brackets with $T_N$, is no longer equivalent to the
grading under which the soliton equations that will be presented in the next
section are homogeneous: $Z,~U_{1,2},~A_2$ now have spin 1, $A_{1,3},~B_{1,2}$
spin $3/2$ and $T_N$ spin $2$.

Actually, the twisted form of the algebra can be written in a rather compact
way using the following notation:
$$\eqalignno{
2 J_{11} = & - 2 J_{22} = U_1 - U_2 ~,~ J_{12} = A_2 ~,~ J_{21} = Z ~,~
U = U_1 + U_2~, \cr
G_1^- = & A_1 ~, G_1^+ = B_2 ~, G_2^- = B_1 ~, G_2^+ = A_3~. &(4.5)\cr
\noalign{\hbox{One finds}}
\{ U,U \} = & - \d_t ~,~ \{ U , G_a^{\pm} \} = \pm G_a^{\pm}~, \cr
\{ J_{ab},J_{cd} \} = & ( \d_{cb} J_{ad} - \d_{ad} J_{cb} ) \d - ( \d_{ad}
\d_{cb} - \ha \d_{ab} \d_{cd} ) \d_t~, \cr
\{ J_{ab},G_c^+ \} = & ( \d_{ab} G_a^+ - \ha  \d_{ab} G_c^+)\d~,\cr
\{ J_{ab},G_c^- \} = & ( -\d_{ac} G_b^- + \ha  \d_{ab} G_c^-)\d~,\cr
\{ G_a^-,G_b^+ \} = & 2 J_{ab} \d_t + ( J_{{ab}_t} + U J_{ab} ) \d + \cr
& \quad \d_{ab} [
\d_{tt} + 2 U \d_t + ( U_t + T_N + 2 J_{ac} J_{cb} + \frac 32 U^2) \d ]~. &
(4.6)\cr} $$
The brackets with $T_N$ are those of primary fields with spins given above, and
for $T_N$ the central term is now $\d'''/2$. All other brackets vanish. One
sees
that the algebra in the above form contains an sl(2) and a u(1) Kac-Moody
algebras. The spin 3/2 fields have a definite u(1) charge and they transform in
the defining representation of sl(2).

One thus recovers a particular example of
the algebras constructed in \BTV\ref\Fuchs{J. Fuchs, Phys.Lett. {\bf B262}
(1991)  249}\ref\Romans{L.J. Romans, Nucl. Phys.{\bf B357} (1991) 549.}. In
\Romans, it was obtained from the standard $u(N-2)$ superconformal algebra
\ref\SCA{V.G.Knizhnik, Theor. Math. Phys. {\bf 66} (1986) 68; M. Bershadsky,
Phys.Lett. {\bf b174} (1986) 285; P.Mathieu, Phys.Lett. {\bf B218} (1989) 185.}
by changing the statistics of the fermionic fields (the resulting algebras
were
called quasi-superconformal). The present case corresponds to $N=4$. On the
other
hand, in \BTV \Fuchs, the general structure was inferred by considering the
embeddings of sl(2) in sl($N$) associated with the decomposition
$\underline{N} \rightarrow \underline2 + (N-2) \underline1$ of the defining
representation.

It is natural to ask whether there is a KdV-type hierarchy related directly to
the quasi-superconformal algebras. By a direct relation we mean a hierarchy
homogeneous with respect to the grading fixed by the quasi-superconformal
algebra (4.6). In fact there exists such a hierarchy: this is exactly the
bosonic
version of the u($N$) super KdV hierarchy introduced in the third reference of
\SCA\ and which we will discuss in more details elsewhere. We just mention that
its first Hamiltonian structure is deduced from the second one by the shift
$T_N \rightarrow T_N + \l $, so that it reads
$$\{ T_N,T_N \} = 2 \d' \quad,\quad \{ G_i^-,G_i^+ \} = \d~~. \eqno(4.7)
$$
In particular, the u(1) quasi-super KdV hierarchy corresponds to the hierarchy
constructed in \Pii\ref\VD{P. van Driel, Preprint IFTA-91-22.} starting from a
gradation intermediate between the principal and the homogeneous ones.

\subsec{The sl$(4)_3$ flows by reduction of the self-dual Yang--Mills
equations.}

Following the method of \ref\MSBD{L.J. Mason and G.A.J. Sparling, Phys. Lett.
{\bf A137} (1989)
 29; I. Bakas and D.A. Depireux, Mod. Phys. Lett. {\bf A6} (1991),
399; L.J. Mason, Twist. Newslett. {\bf 30} (1990) 14.}\ref\Origin{I. Bakas
and D.A. Depireux, Maryland preprint, UMD-PP91-111 (1990), to appear in Int. J.
Mod. Phys. A.} (see also \ref\NSc{V.P. Nair and
J. Schiff, Columbia preprint, CU-TP-521 (1991); V.P. Nair, ``{\it
K\"ahler-Chern-Simons theory}'', University of Columbia preprint, 1991.}), we
can use a  reduction of self-dual Yang-Mills equations in 4-dim to obtain the
fractional KdV equations corresponding to \W43. To this end we start from a
four
dimensional space with signature (2,2) and metric $ds^2 = 2 dx\,dy + 2 dz\,dt$.
With $\epsilon_{xyzt} = -1$, the self-duality equations
$$\left\{ \eqalign{ [D_x,D_t] =&  0 \cr
[D_x,D_y]=& [D_z,D_t] \cr [D_y,D_z] =& 0 \cr} \right.\ \hbox{become} \
\left\{\eqalign{ &[ \pa_t+Q,\pa_x+H] = 0 \cr &[\pa_t + Q, P] = [B,\pa_x+H] \cr
&[P,B] = 0 \cr } \right.\eqno(4.8)
$$
where we have performed a reduction with respect to the two null
Killing symmetries $\pa_y$ and $\pa_z$.

For the matrix $Q$ we take (4.3), whereas for $B$,
we take  $$B = \pmatrix{ 0 & 0 & 0 & 0 \cr
1 & 0 & 0 & 0 \cr
0 & 1 & 0 & 0 \cr
Z & 0 & 1 & 0 \cr } \eqno(4.9)
$$
The reason for this choice will be given section {\it 4.4\/}.

Let us denote the
matrix elements of $H$ by $h_{ij}$. We get, by consistency of (4.8), $h_{14t} =
h_{24t} = h_{34t} = 0 $ and
$$\eqalignno{
h_{13} & = h_{24} \quad,\quad h_{12} = h_{34} \quad,\quad h_{23} = h_{34} -
           h_{14} Z \quad,\cr
h_{11} & = h_{24} Z \quad,\quad h_{22} = - h_{14} U_1 - h_{24} Z \quad,\quad
           h_{33} = - h_{14} U_2 - h_{24} Z \quad,\cr
h_{21} & = - h_{14} A_1 - h_{24} ( 2U_1 + U_2 ) - h_{34} Z     \quad, \cr
h_{32} & = - h_{14} A_2 - 2 h_{24} (  U_1 + U_2 ) - 2 h_{34} Z \quad, \cr
h_{43} & = - h_{14} A_3 - h_{24} U_2 - h_{34} Z \quad, \cr
h_{31} & = - h_{14} B_1 - h_{24} A_2 - h_{34} (U_1 + 2U_2) \quad, \cr
h_{42} & = - h_{14} B_2 - h_{24} A_2 - h_{34} U_1          \quad, \cr
h_{41} & = - h_{14} T  + h_{24} ( Z_t - B_1) - h_{34} A_1 \quad. & (4.10) \cr}
$$

The form of $P$ is given implicitly at the end of section {\it 4.4\/}.
For the fields we find the evolutions
 $$\eqalignno{
Z_x = & - h_{14} Z_t + h_{24} (A_1 - A_3) + h_{34} (U_1-U_2) \cr
U_{1x} = & - h_{14} U_{1t} + h_{24} ( - Z_t - 2 (U_1 + U_2) Z - B_2 ) + h_{34}
(A_1 - A_2 - 2 Z^2) \cr
U_{2x} = & - h_{14} U_{2t} + h_{24} (- Z_t + 2 (U_1+U_2) Z + B_1) + h_{34} (
A_2
- A_3 + 2 Z^2) \cr
A_{1x} = & - h_{14} A_{1t} + \cr
& ~ h_{24} ( -\frac43 U_{1t} - \frac23 U_{2t} + 2
(A_1 - A_2) Z - 3 U_1^2 - 2 U_1 U_2 - U_2^2 - T_0) + \cr
& ~ h_{34} ( -Z_t - 2 (U_1 + U_2)  Z - B_1) \cr
A_{2x} = & - h_{14} A_{2t} +  h_{24} ( - 2 U_{1t} - 2 U_{2t} - 2 U_2^2 +
2 U_1^2) + \cr
& ~ h_{34} ( - 2 Z_t + B_1 - B_2 - 2 (U_2 - U_1) Z) \cr
A_{3x} = & - h_{14} A_{3t} + \cr
& ~ h_{24} (- \frac23 U_{1t} - \frac43 U_{2t} + 2
(A_2 - A_3 ) Z + U_1^2 + 2 U_1 U_2 + 3 U_2^2 + T_0 )  + \cr
& ~ h_{34} ( - Z_t + B_2 + 2  (U_1 + U_2) Z ) \cr
B_{1x} = & - h_{14} B_{1t} + h_{24} ( - A_{2t} + 2 Z B_1 + 2 (U_1 + U_2)
(A_1-A_2)\cr
& + h_{34} ( - \ha U_{1t} - \frac32 U_{2t} + 2 (A_1 - A_2) Z - U_1^2 - 2 U_1
U_2 - 3 U_2^2 - T_0)\cr
B_{2x} = & - h_{14} B_{2t} + h_{24} ( - A_{2t} - 2 Z B_2 + 2 (U_1 + U_2)
(A_2 - A_3) \cr
& + h_{34} ( - \frac32 U_{1t} - \ha U_{2t} + 2 (A_2 - A_3) Z + 3 U_1^2 + 2
U_1 U_2 + U_2^2 + T_0)\cr
T_{0x} = & - h_{14} T_{0t} + h_{24} (- \frac23 (B_1 + B_2) + \frac43 (U_1 +
U_2) Z )_t \cr
& + h_{34} ( - \frac13 (A_1 + A_2 + A_3) + \frac13 Z^2)_t & (4.11) \cr}
$$

Inspection of the equations shows that the spins of $h_{14}$, $h_{24}$ and
$h_{34}$ differ in ascending order by $1/3$. If all three coefficients are
zero,
 all the flows are trivial. If we set $h_{14}=1$, then since $h_{24}$ and
$h_{34}$ have spins $1/3$ and $2/3$ resp. and they are constant, we must set
them
equal to zero. For the same reason,  the only other two solutions are ($h_{14}
=
0$, $h_{24} = 1$, $h_{34} = 0 $) and ($h_{14} = 0$, $h_{24} = 0$, $h_{34} = 1
$).
Each different solution gives rise to a set of equations which are Hamiltonian.
Explicitly, we have $$\matrix {
h_{14} = 1 ~,& H_{1\phantom{/3}} = \int - T_0  dt                \hfill \cr
\noalign{\vskip3pt}
h_{24} = 1 ~,& H_{2/3} = \int [-(B_1 + B_2) + 2 Z (U_1 + U_2)]dt \hfill \cr
\noalign{\vskip3pt}
h_{34} = 1 ~,& H_{1/3} = \int [- (A_1 + A_2 + A_3) + Z^2]dt      \hfill \cr
  } \eqno(4.12)
$$
Note that the conformal dimension of these Hamiltonians is well defined.
If we had modified the energy-momentum tensor, as in the previous section, in
order to get a conformal algebra with all fields primary, we would have found
$H_{1/3} = \int - (A_1 + A_2 + A_3) + \ldots$, but since the fields $A_i$'s
have
different spins, the Hamiltonian is not dimensionally homogeneous.

\subsec{Relation with the results obtained by $x\leftrightarrow t$
interchange.}

As already stated, the \W43 algebra obtained by $x \leftrightarrow t$
interchange and the one obtained by Hamiltonian reduction are fully equivalent.
The fields in the two approaches are related by $$\eqalignno{ 4 Z =& u_2
\quad,\quad 8 U_1 = - v_2 + 2 u_3 \quad,\quad 8 U_2 = - 3 v_2 + 2 u_3 \quad,
\cr
16 A_1 =& 16 u_4 - 2 w_2 - 4 v_3  - u_2{}^2 \quad,\quad
16 A_2 = 16 u_4 - 8 v_3 + 3 u_2{}^2 \cr
16 A_3 =& 16 u_4 + 6 w_2 - 12 v_3  - u_2{}^2 \cr
4 B_1 =& - 3 u_{2t} + 4 w_3 - 10 v_4 + 3 u_2 v_2 - u_2 u_3 \cr
4 B_2 =& - 5 u_{2t} + 6 w_3 - 14 v_4 + 5 u_2 v_2 - u_2 u_3 \cr
T_0 =& \frac1{10} w_4 + \frac 7{40} v_{2t} - \frac1{10} u_3^2 - \frac{61}{160}
v_2^2 + \frac7{80} u_2 w_2 + \frac3{40} u_2 v_3 + \cr
& \quad \frac{33}{40} v_2 u_3 +
\frac1{16} u_2{}^3 - \frac38 u_3{}^2. & (4.13) \cr }
$$
With these field redefinitions and $t \rightarrow -t$, the flow associated to
$H_{1/3}$ is easily checked to be equivalent to the first flow of the sl$(4)_3$
KdV hierarchy (3.21).
At
this point, we recall that when regarding the $W_4$ algebra as a second
Hamiltonian structure, we obtain the first Hamiltonian structure by shifting
$u_4$ by a constant. Inspection of the field redefinitions shows that such a
shift corresponds to shifting each of the $A_i$ fields by the same constant. So
the first Hamiltonian structure for the sl$(4)_3$ flows can be obtained by such
a
shift, and then the usual procedure can be employed to generate an infinite
hierarchy of flows and conserved quantities, recovering the integrability
properties from another point of view.

Also, in \me{} it was noticed that
the same shift relates $Q$ to $B$, and $H$ to $P$, in the following sense: if
the first and the second  Hamiltonian structures are related by a shift of,
say,
the fields $q$, $\{.,.\}_{2,q+\l} =   \{.,.\}_{2,q} + \l \{.,.\}_{1,q}$, then
$Q(q+\l) = Q(q) + \l B$ and  $H(q+\l) = H(q) - \l P$. Here we should remember
that the field $T$ which  enters
$Q$ is related to the energy-momentum tensor $T_0$ by $T = T_0 + A_2 Z +
\ldots$
Therefore, we can see that a similar relation exists here between $Q$ and $B$.
We observed that the same relation holds true for $H$ and $P$. This remark
gives a heuristic justification for the form of the $B$ matrix we took in
(4.9).

\newsec{On the general structure of \W Nl algebras.}

In this section we want to present some general characteristics of the \W Nl
algebras which can be extracted from the $x\leftrightarrow t$ interchange and
the Hamitonian reduction methods.

\subsec{Canonical basis of independent fields for the sl$(N)_l$ hierarchy and
spin content of the \W Nl algebra}

In this section, $N$ and $l$ will be taken to be coprime. To perform the
$x\leftrightarrow t$ interchange in the $l^{th}$ flow of the sl($N$) KdV
hierarchy, one has to introduce a certain number of new independent fields
which are the first few derivatives of the sl($N$) KdV fields $u_2$, $u_3$,
$\ldots u_N$. This set must be chosen such that the $x$ derivative of every
field is either another field of the set or can be expressed in terms of the
time
derivative of other fields of the set using the $l^{th}$ sl($N$) KdV equation.
A
convenient basis for these new independent fields is given by
$$
\{ u_i,u_{ix},u_{ixx},\ldots,u_i{}^{(l-1)} \} ~,i = 2,\ldots N.\eqno(5.1)$$
This will be called the canonical basis. It consists of $l(N-1)$ fields. From
such a basis one can read off directly the spin content of the fields in the \W
Nl algebra. These are  $$
{\rm spins\ }W_N{}^{(l)} = \{ {i+k \over l}, i=2,\ldots,N; k=0,\ldots,l-1 \}
\eqno(5.2)$$
(Recall that spin $u_i{}^{(k)} = (i+k)/l$.) This spin content satisfies the sum
rule (as conjectured in \CUNY)
$$
\sum_{{\rm spins}\, s} (2s-1) = \sum_{i=2}^{N} \sum_{k=0}^{l-1} (2 {(i+k) \over
l} - 1)  = N^{2} - 1 =  {\rm dim} ~ {\rm sl}(N)~~.\eqno(5.3)$$
The basis (5.1) is of course not unique but any other basis has exactly the
same spin content. For instance, in the sl(3) case, one can write the second
flow under the form
$$\eqalign{u_{2t} & = - u_{2xx} + u_{3x}   \cr
           u_{3t} & = u_{3xx} + u_2 u_{2x} \cr } \eqno(5.4) $$
The canonical basis corresponds to the choice of the new fields
$$v_2 = u_{2x} \quad,\quad v_3 = u_{3x} \eqno(5.5)$$
but one could also have considered
$$v_2 = u_{2x} \quad,\quad w_2 = u_{2xx} \eqno(5.6)$$
However these two choices yield equivalent sl$(3)_2$ hierarchies and in
particular $\tilde P_2$ is the same for both cases, up to simple field
redefinitions. This generalized to more complicated cases.

\subsec{What happens when $(l,N) \neq 1$?}

When $l$ and $N$ are coprime, it always appear to be possible to construct the
sl$(N)_l$ hierarchy by direct $x \leftrightarrow t$ interchange.
However, this is not the case when $(l,N) \neq 1$. For instance, consider the
second flow of the sl(4) hierarchy:
$$\eqalign{
u_{2t} = & - 2 u_{2xx} + 2 u_{3x} ~,\cr
u_{3t} = & - 2 u_{2xxx} + u_{3xx} + 2 u_{4x} - u_2 u_{2x} ~,\cr
u_{4t} = & - \ha u_{2xxxx} + u_{4xx} - \ha u_2 u_{2xx} - \ha u_3 u_{2x} ~,\cr }
\eqno(5.7) $$
It turns out here that for any choice of independent fields, the $x$-derivative
of one of the fields is not determined by the above equation. This problem
persists at the level of the modified equations. The Hamiltonian for the flow
(5.7) is simply $\int u_3 dx$. When reexpressed in terms of modified fields,
one
easily gets the modified equations
$$\eqalign{
r_{1t} = & r_{3xx} - (r_1 r_2)_x ~, \cr
r_{2t} = & - \ha (r_1^2 - r_3^2)_x ~, \cr
r_{3t} = & - r_{1xx} + ( r_2 r_3)_x ~.\cr} \eqno(5.8) $$
Therefore if we choose the new independent fields to be $s_i = r_{ix}$, one
sees that $s_{2x}$ is not determined by the above equations due to the absence
of a term $r_{2xx}$.

It is simple to show that this situation is generic for $l=2$ and $N$ even.
Indeed, up to non-linear terms, the evolution equation for $u_i$ reads
$$ u_{it} = u_{ixx} + 2 u_{i+1~x} - \frac 2N {N\choose i} u_2{}^{(i)} + \ldots
\eqno(5.9) $$
where ${N\choose i} $ denotes a binomial coefficient. Let us fix the basis to
be \hfil\break $\{u_2,u_{2x},\ldots,u_2{}^{(N-1)},u_3,u_4,\ldots,u_N \}$, in
terms of which the argument is simpler. We want to show that the $x$-derivative
of $u_2{}^{(N-1)}$ is not determined by (5.7). From the $N-3$ first equations
in
(5.7), one expresses $u_{kx}$ $(k\neq2)$ in terms of the other fields of the
above set:
$$ u_{kx} = \frac1N u_2{}^{(k-1)} \sum_{i=1}^{k-1} (-)^{i+1} {N\choose
{N-i}} \frac 1{2^{i-1}}
\eqno(5.10)$$
In the final equation, the coefficient of $u_2{}^{(N)}$ is
$$\frac1n \sum_{i=1}^{N-1} {N\choose i} \frac {(-)^{N-i+1}}{2^{N-i+1}} -
\frac2N
= \frac{(-)^N - 1 }{N\ 2^{N-1}}\eqno(5.11) $$
This vanishes when $N$ is even, in which case $u_2{}^{(N)}$ is not determined.
The argument works for other choices of the basis.
We expect this to be generic to the cases where $N$ and $l$ are not coprime,
but we haven't found a direct check of this statement within the above
approach.

\subsec{A comment on the Hamiltonian structure of the modified fields for the
sl$(N)_l$ hierarchy.}

Let us introduce the following basis for the modified fields:
$$\{ \f_i{}^{(k/l)} ,i=1,\ldots,N-1;k=1,\ldots,l-1 \} \eqno(5.12)$$
where $\f_i{}^{(1/l)}$ are the usual modified fields $r_i$ introduced
previously,
up to a possible scaling factor, and $\f_i{}^{(k/l)}$ is linearly related to
$r_i{}^{(k+1)}$. Notice that the new dimension of $\f_i{}^{(k/l)}$ is
just $k/l$. From the first few examples which have been worked out, it is
tempting to guess that for suitably chosen $\f_i{}^{(k/l)}$, the Hamiltonian
structure of the modified fields will read
$$\eqalign{
\{ \f_i{}^{(k/l)} , \f_j{}^{((l-k)/l)} \} & = - \d_{ij} \d \qquad k \neq l/2
\cr
\{ \f_i{}^{(1/2)} , \f_j{}^{(1/2)} \} & = - \d_{i,N-i-1} \d  \cr
\{ \f_i{}^{(1)} , \f_j{}^{(1)} \} & = - \d_{ij} \d_t \cr } \eqno(5.13)$$
This structure is certainly ill defined for $N$ and $l$ even.
For instance, for \W42, this would give $\{r_2,r_2\} = - \d$, which is
impossible.

\subsec{Results from the Hamiltonian reduction}

For the $W_N$-algebras (i.e. \W N1 in our notation), the matrix $A(t)$
corresponding to (4.1) is constrained by setting the first upper diagonal to
$-1$. It can always be brought by Hamiltonian
reduction to the form of a matrix with zero entries everywhere except for the
first upper diagonal which is set to $-1$ and the lowest row which takes the
form  $$\left\lfloor\matrix{
0   &   0   &  0  &  -1   \cr
u_N &\ldots & u_2 & 0     \cr}\right\rfloor \eqno(5.14)$$
There are field redefinitions so that the corresponding $W$-algebra
is conformal. It is well-known that he energy-momentum tensor takes the form
$$T_0 = u_2
= - \ha tr J^2 + \frac 1l \pa ((n-1) J_{11} + (n-2) J_{22} + \ldots +
J_{n-1,n-1} ) \eqno(5.15)
$$
with $l=1$.

For the \W N2 algebras, we performed the Hamiltonian reductions up to $N=8$ and
found the following results. The constraint we impose is to set the second
upper diagonal to $-1$. This constraint is preserved by gauge transformations
generated by strictly lower triangular matrices. However, for $N$ even, it is
not possible to fix this gauge freedom completely in a local way. So when
considering gauge transformations of the form (4.2), we set $g_{21} = 0$ if $N$
is even. The hamiltonian reduction gives the following:
the ``$Q$'' matrix can always be reduced so that it is zero everywhere, except
for the second upper diagonal which is $-1$, and the lower two rows. These rows
appear as follows:
$$\left\lfloor
\matrix{ 0 & \ldots & 0   & 0     &    0    & -1    \cr
         * & \ldots & T_1 & G^{(+)} &  U    &  Z    \cr
         * & \ldots & *   & T_2     & G^{(-)} &  -U \cr
}\right\rfloor
\eqno(5.16)$$
The field $Z$ is identically zero if $N$ is odd.
We find that $T_0 = T_1 + T_2 - U^2 - Z G^{(-)} + \ha
U'$ is an energy-momentum tensor,  whose expression in terms of the original
fields is (5.15) with $l=2$ modulo terms involving derivatives of the fields
above the diagonal.
With respect to this
energy-momentum tensor, $Z$ has spin $1/2$,  $U$ is quasi-primary of spin 1,
etc. $T_0$ is {\it the\/} energy-momentum that respects the original grading of
the KdV fields, as given by the $ x \leftrightarrow t $ interchange.

The presence of the $Z$ field for $N$ even is a reflection, in the context of
Hamiltonian reduction, of the presence of constraints in the approach where $x$
and $t$ are interchanged. From this latter point of view, since the $Z$ field
has spin 1/2, that means that it originally had spin 1 before the interchange.
Since there is no spin 1 field in the sl(4) KdV hierarchy, this indicates that
such a field has to be introduced to take care of the constraints.

For the more general \W Nl algebras, let us know restrict ourselves
for convenience to the cases where $N$ and $l$ are coprime. For the \W N3
cases,
we find by Hamiltonian reduction a ``$Q$'' matrix that has zeros everywhere but
for the third upper diagonal set to -1 and the lowest three rows, in which we
find one spin-2/3 field, two spin-2 fields, $\ldots$ They are arranged as
follows:

$$\left\lfloor
\matrix{ 0 & \ldots &  0  &  0   &  0  &   0  &  0  & -1          \cr
         0 & \ldots & T_1 & B_1  & A_1 &  U_1 &  Z  &  0          \cr
         * & \ldots &  *  & T_2  & B_2 &  A_2 & U_2 &  0          \cr
         * & \ldots &  *  &   *  & T_3 &  B_3 & A_3 & - U_1 - U_2 \cr
}\right\rfloor \eqno(5.17)$$
for the \W {{3n+1}}3, whereas for the \W {{3n+2}}3, the spin $2/3$ field $Z$ is
in position $(N-1,N)$.
The combination $T_0 = T_1 + T_2 + T_3 - U_1^2 - U_2^2 - U_1 U_2 - A_i Z +
\frac 23 U_1' + \frac13 U_2'$ (with $i=2$ for \W {{3n+1}}3 and i=3 for \W
{{3n+2}}3) forms the energy-momentum tensor. The explicit expression of
this tensor in terms of the original currents is
again (5.15) with $l=3$ and terms involving derivatives of the fields above
the diagonal.

It is clear from the above examples and their natural extension that for $N$
and
$l$ coprime, Hamiltonian reduction leads to a \W Nl spectrum that corresponds
identically to the one dictated by the $x \leftrightarrow t$ interchange. In
addition, the energy-momentum tensor that respects the natural grading of the
KdV equations takes the form
$$ \eqalign{ T_0(W_N{}^{(l)}) = - \ha tr J^2 -
\frac1l (\sum_1^{N-1} & (N-i) J_{i,i}') + \cr
& ({\rm derivatives\ of\ non-diagonal\ elements}) \cr}\eqno(5.18)
$$
We conclude that $\{T_0,T_0\} = \frac{(N^3-N)}{12 l^2} \d_{ttt} + 2 T_0 \d_t +
T_{0t} \d$.

\newsec{Conclusions.}

In this work we have constructed a new explicit example of a \W Nl algebra,
\W 43, from the point of view of generalized KdV hierarchies
(\MO,\me,\Pi,\Pii). We have thus provided further support to the conjecture
that the \W Nl algebras correspond to the second Hamiltonian structure of the
sl$(N)_l$ KdV hierarchy. The latter refers to the new hierarchy obtained from
the standard sl($N$) KdV hierarchy by interchanging the roles of the variables
$x$ and $t$ in the $l^{th}$ flow. Granted this conjecture, another original
motivation was to advocate this approach as being a simple and systematic way
of
constructing the \W Nl algebras. From this point of view, the present analysis
shows that it is not as simple as initially expected. At first we found from
the
outset that this approach works straightforwardly only when $N$ and $l$ are
coprime. We plan to return to the cases $(N,l)\neq 1$ elsewhere, but these are
certainly more complicated since one has to deal with constrained systems.
Second, for the new example we have worked out, the expression we obtain for
this second Hamiltonian structure is quite complicated. This Poisson structure
can be somewhat simplified by introducing a new set of independent fields,
namely those fields which appear naturally in a zero curvature formulation, or
equivalently, the method of Hamiltonian reduction. If we modify the
energy-momentum tensor to make them primary, then the spin content of the
algebra is modified. At this step, it appears that the algebra acquires a much
nicer form once we recognize the existence of an underlying Kac-Moody algebra
organizing the whole conformal algebra. In fact we recover an example of a $W$
algebra one obtains by considering a non-principal embedding of sl(2) (the
sum-embedding) into sl(N) \BTV,  or equivalently of a u($N-2$) quasi-conformal
superalgebra \Romans. We point out that the latter is also related to the
bosonic form of the u($N-2$) super KdV hierarchy. It is certainly interesting
and satisfying to display explicitly the perfect equivalence of the method of
$x
\leftrightarrow t$ interchange and that of non-principal sl(2) embeddings,
which
from the outset looks conceptually rather remote.

Finally let us emphasize some favorable aspects of the method of $x
\leftrightarrow t$ interchange. That we derive the second Hamiltonian structure
via the modified fields gives us in one shot both the algebra and its free
field representation. Also the method emphasizes the fact that everything we
want
to know about \W Nl and the sl$(N)_l$ hierarchy can be extracted
systematically from $W_N$ and the usual sl($N$) hierarchy. Hence, although
there
is no direct relation between the various \W Nl algebras for $N$ fixed, the
method unravels one such hidden relationship: all these algebras have the same
Miura transformation, which is however written differently for different values
of $l$.

\vskip.5in
\centerline{\bf Acknowledgements}

We would like to thank I.Bakas for
a lot of help in the early stage of this project and subsequent discussions,
and P. van Driel for discussions and comments. This work was supported by NSERC
(Canada), FCAR (Qu\'ebec) and BSR (U.Laval).

\listrefs

\end